\newlength{\extralineskip}
\documentclass[12pt]{article}
\usepackage{epsfig}

\newcommand{\be}{\begin{equation}}
\newcommand{\ee}{\end{equation}}
\newcommand{\qvec}{\mathbf{Q}}
\begin{document}
\begin{titlepage}
\begin{flushright}
          \begin{minipage}[t]{10em}
          UAB--FT--481\\
                 January 2000
          \end{minipage}
\end{flushright}
\vspace{\fill}

\vspace{\fill}

\begin{center}
\baselineskip=2.5em

{\large \bf Effects of Bose-Einstein Condensation on forces among bodies
sitting in a boson heat bath }
\end{center}

\vspace{\fill}

\begin{center}
{\bf F. Ferrer and J. A. Grifols}\\
\vspace{0.4cm}
     {\em Grup de F\'\i sica Te\`orica and Institut de F\'\i sica
     d'Altes Energies\\
     Universitat Aut\`onoma de Barcelona\\
     08193 Bellaterra, Barcelona, Spain}
\end{center}
\vspace{\fill}

\begin{center}
\large Abstract
\end{center}
\begin{center}
\begin{minipage}[t]{36em}
We explore the consequences of Bose-Einstein condensation on
two-scalar-exchange mediated forces among bodies that sit in a
boson gas. We find that below the condensation temperature the range of the
forces becomes infinite while it is finite at temperatures above condensation.
\end{minipage}
\end{center}

\vspace{\fill}

\end{titlepage}

\clearpage

\addtolength{\baselineskip}{\extralineskip}

 Van der Waals type forces where two photons are being
exchanged~\cite{waals} or the extremely feeble forces generated by
2-neutrino exchange~\cite{neut,sik} provide examples of forces among two
static bodies in a vacuum produced by the exchange of two quanta in the
t-channel. Spin independent interactions arising
from double (pseudo)scalar exchange~\cite{pseudo1,pseudo2,pseudo3} (such as
axions and/or more
bizarre specimens of modern completions of the Standard Model(SM))
provide
further examples for these so-called dispersion
forces~\cite{sucher}. When the objects that feel such forces are placed
in a heat bath
at a temperature T, the forces get modified.
Indeed, in the case of molecules in the relic photon background, the
long-range Casimir-Polder forces among them are
strongly affected for distances much larger than $T^{-1}$~\cite{fgcp}
and, for the
2-neutrino forces, the cosmic neutrino background
completely screens off the interaction at large distances~\cite{fgn}
(again, large
meaning much larger than $T^{-1}$).

In the present paper we shall deal with a gas of scalar bosons carrying an
abelian charge and non-zero chemical potential. As mentioned
before, their double exchange between fermions has been
studied in vacuum. The case of a non-charged scalar bath in a classical
Boltzmann distribution was briefly discussed in~\cite{pseudo3}. However we are not aware of discussions on the
effects resulting from placing the interacting system
in a charged scalar heat bath displaying genuine quantum statistical effects
such as Bose-Einstein condensation. Because in the previously reported instances,
interesting effects did result, we think it is
worthwhile to raise this issue here. Admittedly, light scalar bosons
have a much different status than photons and
neutrinos and their nature is entirely speculative. Nonetheless, in
almost any extension of the Standard Model,
scalars are present and some have been furthermore suggested as
candidates for dark matter so that might be part of
the cosmic relic background. Not to speak about axions needed to solve
the serious CP problem of QCD and sometimes
advertised also as galactic halo dark matter or even as dark matter on a
cosmic scale.

Since we do not pretend to resemble any particular model of matter
scalar interactions we will construct a very
simple toy model that avoids unessential complications and that might
mimic more "realistic" interactions of
matter and light scalar fields. Consider the interaction lagrangean,
\be
{\cal L}_{int}= g \Phi^2 \varphi^2 
\label{eq:lint}
\ee
$\Phi$ is a heavy scalar field of mass $M$ and $\varphi$ is a light
scalar
field of mass $m$. Let us now put two
such heavy particles $\Phi$ in vacuum at a distance $r$. Their lowest
order interaction is given by the
Feynman amplitude in figure 1. The potential is obtained from the NR
limit of this amplitude via Fourier
transformation. That is,
\begin{figure}[bht]
\begin{center}
\epsfig{file=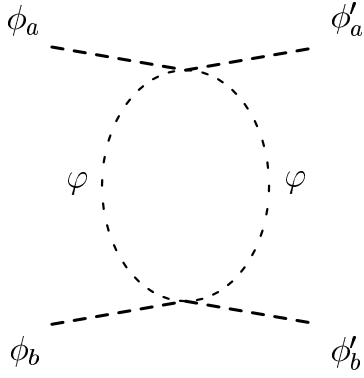,width=5.5cm,height=5cm}
\end{center}
\caption{\it Feynman diagram corresponding to the interaction which gives rise
  to the long-range force.}
\end{figure}
\be
V(r)=i\int{\frac{d^3 \qvec}{(2 \pi)^3} e^{i \qvec \cdot \mathbf{r}}
  \frac{{\cal T} \left(q \simeq (0,\qvec) \right)}{4 M^2}}
\label{eq:defpot}
\ee
where
\be
{\cal T} (q)=g^2 \int{\frac{d^4 k}{(2 \pi)^4} \frac{1}{k^2-m^2+i
    \epsilon}\; \frac{1}{(q-k)^2-m^2+i \epsilon}}.
\label{eq:mnorel}
\ee
Taken at face value the previous integral diverges. It has to be
regulated and the piece which leads to a long-range interaction
extracted. We
are left with a function of $\log Q^2$, with $Q\equiv|{\mathbf{Q}}|$, after
dropping any polynomial in $Q$ that would only yield contact
interactions~\cite{sik,consoli}. 
Introducing
a convergence factor $e^{-\eta Q}$ to evaluate the Fourier transform and
taking the limit $\eta \rightarrow 0$ we obtain:
\be
V_{vac}(r)=-\frac{g^2 m}{64 \pi^3 r^2 M^2} K_1 (2 m r)
\label{eq:bessel}
\ee
for the potential. The same result can be derived using the dispersion
theoretic techniques of Feinberg and Sucher~\cite{sucher}.

For small $\varphi$ mass, eq.~(\ref{eq:bessel}) becomes:
\be
V_{vac}(r)=-\frac{g^2}{128 \pi^3 r^3 M^2}
\label{eq:potbuit}
\ee
valid for $r\ll 1/m$. Beyond this range the Bessel function gives rise
to the
characteristic Yukawa factor $e^{-2 m r}$.

Incidentally, identical $r$ behaviour obtains for the spin independent
potential arising in double exchange of  pseudoscalars
coupled to matter fermions via Yukawa couplings~\cite{pseudo2,pseudo3}.  

Next we introduce our system in a heat reservoir made of an ideal
relativistic $\varphi$ gas at temperature $T$ ($T>m$). We further assume
that the particles in the gas carry a conserved
quantum number (which we will refer to as "charge") corresponding to a
quantum mechanical operator $\cal Q$. We may use real
time finite temperature field theory~\cite{kobes} to calculate the
effect of the heat bath on the potential between the two
massive particles. So, we take for the  $T$-dependent
$\varphi$-propagator,
\be
D_F(k,T)=\frac{1}{k^2-m^2+i \epsilon}-2 \pi i \delta(k^2-m^2)
\left[\theta(k^0) n(|k^0|,T)+\theta(-k^0) \bar{n}(|k^0|,T) \right]
\label{eq:propt}
\ee
where 
\be
n(\omega,T)=\left( e^{(\omega-\mu)/T}-1\right)^{-1} \quad \quad
\mathrm{and}
\quad \quad \bar{n}(\omega,T)=\left( e^{(\omega+\mu)/T}-1\right)^{-1} 
\label{eq:nbose}
\ee
are the B-E distribution functions for particles (charge +1) and
antiparticles (charge -1),
respectively. $\mu$ is the chemical potential associated to the
conserved charge $\cal Q$.

 The amplitude of figure 1 now generalizes to
\be
{\cal T} (q)=g^2 \int{\frac{d^4 k\:d^4 k'}{(2 \pi)^4}
\delta^{(4)}(k+k'-q)
D_F(k,T)
  D_F(k',T)}. 
\label{eq:mgen}
\ee
This amplitude generates two distinct contributions to the potential.
The first one arises from the first piece in
$D_{F}(k,T)$ and is the vacuum potential just derived. The other
corresponds to the situation where one of the
scalars in the double exchange process is supplied by the thermal bath.
This effect is described by the crossed terms in
the amplitude involving the thermal piece of one $\varphi$-propagator
along
with the vacuum piece of the other propagator.
This thermal component of the Feynman amplitude can be written as
\be
{\cal T}_{\:T} (q)=-i g^2 \int{\frac{d^3 \mathbf{k}}{(2
\pi)^3}\frac{1}{2 \sqrt{\mathbf{k}^2+m^2}}
  \left[ \frac{1}{(k-q)^2-m^2} + \frac{1}{(k+q)^2-m^2} \right]
\left(n+\bar{n}
  \right) }
\label{eq:mther}
\ee
where use has been made of energy-momentum conservation and of the
$\delta (k^2-m^2)$ in $D_{F}(k,T)$. In the static limit,
 i.e. momentum transfer
$q\simeq(0,\qvec)$, where matter is supposed to be at rest in the frame
of the heat reservoir, the piece in square
brackets in eq.~(\ref{eq:mther}) reads:
\be
-2 \left(Q^2-4 \mathbf{k}^2 \hat{\qvec} \cdot \hat{\mathbf{k}}
\right)^{-1}.
\label{eq:simpl}
\ee
So, finally the $T$-dependent amplitude to be
Fourier transformed is
\be
{\cal T}_{\:T} \left( q\simeq(0,\qvec)\right)=i g^2
\int{\frac{d^3\mathbf{k}}{(2
    \pi)^3}\frac{1}{\sqrt{\mathbf{k}^2+m^2}}\frac{1}{Q^2-4 \mathbf{k}^2
    \hat{\qvec} \cdot \hat{\mathbf{k}}}\left(n+\bar{n} \right)}
\label{eq:beft}
\ee
i.e. it has been reduced to an integral over the phase space of the real
particles (and antiparticles) in the heat bath. 

The reservoir is thermodynamically characterised by a temperature $T$, a
volume $\cal V$, and a fixed charge $\cal Q$\footnote{For definiteness
we take ${\cal Q}>0 $, i.e. particles outnumber
antiparticles.}. Then, the chemical potential $\mu(T)$ is determined from
the relation
\be
{\cal Q}=\sum_{k}\left(n-\bar{n}\right).
\label{eq:defmu}
\ee
For a Bose-Einstein gas, the sum over states in the previous formula can
be converted to an integral like the one in eq.~(\ref{eq:beft})
as long as its temperature is above a critical temperature $T_{c}$.
Below that temperature, if one makes the replacement
\be
\sum_{k}\mapsto {\cal V} \int{\frac{ d^{3}\mathbf{k}}{(2\pi)^3}}
\label{eq:sumint}
\ee
in eq.~(\ref{eq:defmu}), the result is less than $\cal Q$~\cite{haber}.
This is
because below $T_{c}$ a
large macroscopic fraction of the charge resides in
the lowest energy state and the density of states ${\cal V}k^2/2\pi^2$
in the
continuous representation of the sum over states
gives a zero weight to the zero mode. On the contrary if the gas is
above $T_{c}$ then the charge is thinly distributed
over the states and no individual state is populated by a macroscopic
fraction of the total charge so that by passing to
the continuum essentially only an infinitesimally small error is done.
Let us discuss both cases in turn. First start with
the non degenerate case, i.e. when $T$ is above the condensation
temperature $T_{c}$. In this instance, the phase space
integrals in the formulas above  correctly describe the physics of the
problem. Therefore we can take eq.~(\ref{eq:beft}) and plug it in
the expression for the potential~(\ref{eq:defpot})
\be
V_T(r)=-\frac{1}{2 \pi^2}\frac{1}{r}\frac{g^2}{4 M^2}\int_0^{\infty}{d Q
  \:Q\:\sin{Q r}\int{\frac{d^3 \mathbf{k}}{(2
\pi)^3}\frac{1}{\sqrt{\mathbf{k}^2+m^2}}
      \frac{1}{Q^2-4 \mathbf{k}^2 \hat{\qvec} \cdot \hat{\mathbf{k}}}
\left( n+\bar{n}
\right)}}.
\label{eq:plugin}
\ee
After trivial integration over $Q$ and over the polar angle in
$\mathbf{k}$-space
we get
\be
V_T(r)=-\frac{g^2}{64 M^2 \pi^3} \frac{1}{r^2}
\int_0^{\infty}{\frac{\mathbf{k}^2
    d|\mathbf{k}|}{\sqrt{\mathbf{k}^2+m^2}}
  \frac{\sin{2
|\mathbf{k}|r}}{|\mathbf{k}|}\left[\frac{1}{e^{\left(\sqrt{\mathbf{k}^2+
m^2}-\mu
\right)/T}
      -1}+\frac{1}{e^{\left(\sqrt{\mathbf{k}^2+m^2}+\mu \right)/T} -1}
\right]}.
\label{eq:pother}
\ee
This equation by itself is not sufficient to determine $V_{T}(r)$ for
the functions $n$ and $\bar{n}$ contain the chemical
potential $\mu(T)$ which has to be obtained through
\be
\rho \equiv \frac{\cal Q}{\cal V}=\frac{1}{2 \pi^2} \int{\mathbf{k}^2
d|\mathbf{k}| \left(n-
\bar{n}
  \right)} 
\label{eq:defrho}
\ee
which, by the way, also determines the critical temperature $T_{c}$ via
the implicit equation~\cite{haber}:
\be
\rho=\rho \left(T=T_c,\mu=m\right).
\label{eq:deftc}
\ee
So eqs.~(\ref{eq:pother})~and~(\ref{eq:defrho}) give the solution to our
problem. We can use the high-temperature expansion of eq.~(\ref{eq:defrho}) derived in~\cite{haber} to obtain the
chemical potential as a function of $T$. To leading order,
$\mu(T)=m(T_c/T)^2$, and we introduce it
in eq.~(\ref{eq:pother}) to get our potential above the condensation temperature. The
integral in eq.~(\ref{eq:pother}) can be easily
split in three pieces upon using the following identity:
\be
{1\over e^{y}-1}={1\over y}-{1\over 2}+2\sum _{k=1}^{\infty}{y\over
y^{2}+(2\pi k)^{2}}
\label{eq:ident}
\ee
The resulting integrals can be found in~\cite{gradstein}. The $-{1\over 2}$
piece exactly cancels the vacuum 
potential~(\ref{eq:bessel}) so that the total contribution to the potential is
finally:
\be
V_{T\ge T_c}^{total}=-\frac{g^2}{64 M^2 \pi^2}
\frac{1}{r^2}T\left[e^{-2mr\sqrt {1-\xi^4}}+2\sum_{k=1}^\infty e^{-
 rT\sqrt{2(\alpha +\beta)}}\cos \left(rT\sqrt{2(\alpha-\beta}\right)\right]
\label{eq:vtotmes}
\ee
with $\alpha\equiv \sqrt{(4k\pi
mT_{c}^2/T^3)^2+[m^2(1-\xi^4)/T^2+4k^2\pi^2]^2}$, $ \beta\equiv
4k^2\pi^2+
m^2(1-\xi^4)/T^2$, and $\xi \equiv T_c/T$. 
 
Notice in the first term of eq.~(\ref{eq:vtotmes}) the typical Yukawa damping factor
that cuts off the interaction at distances 
long compared to the Compton wavelength of $\varphi$. Furthermore, since
for any $k$ 
\be
T\sqrt{2(\alpha+\beta)}>2m\sqrt{1-\xi^4}
\ee
all modes in the second term of eq.~(\ref{eq:vtotmes}) are even more suppressed at large
distances. This makes sense physically and is very
convenient when trying to do a numerical evaluation.

Below $T_{c}$, a macroscopic fraction of the charge carried by particles
in the reservoir piles up in the zero mode state
(the condensate) and the integrals in
eqs.~(\ref{eq:pother})~and~(\ref{eq:defrho}) no longer correctly
describe the physical situation. Indeed, eq.~(\ref{eq:defrho}) gives
the density of charge in excited states $\rho^*$, i.e. the thermal
modes~\cite{haber}. For a relativistic boson gas,
\be
\rho^*=\frac{1}{3} m T^2
\label{eq:rhoexc}
\ee
where we used $\mu=m$ in this temperature regime since $\mu$ has to be
always less or equal than $m$ and it monotonically
increases as the temperature decreases until it reaches $m$ at $T_{c}$
(and stays fixed in the macroscopical sense
thereafter). Because the definition of the condensation temperature
eq.~(\ref{eq:deftc}) implies in this case
\be
\rho=\frac{1}{3} m T^2_c\, ,
\label{eq:rhotot}
\ee
the charge density in the ground state is
\be
\rho_0=\rho \left(1-\left(\frac{T}{T_c}\right)^2 \right).
\label{eq:rhocond}
\ee
The Feynman amplitude eq.~(\ref{eq:beft}), which involves also a sum
over
states, should be split accordingly in two parts. The zero mode term on
the one hand and on the other hand the integral over thermal modes. The
zero momentum mode contributes to the amplitude
\be
\left. {\cal T}_T \left( Q \right) \right|_{k=0}=\frac{i
g^2}{m Q^2}\frac{1}{\cal V}
\left. \left( n+\bar{n}\right) \right|_{k=0}
\label{eq:mcond}
\ee
where the distribution function factor can be rewritten as
\be
\frac{1}{\cal V}\left. \left( n+\bar{n}\right) \right|_{k=0}=\frac{\cal
Q}{\cal
V}\left(1-\left(\frac{T}{T_c}\right)^2 \right)+\frac{1}{\cal
V}\frac{2}{e^{2m/T}-1}
\label{eq:ncond}
\ee
since $\mu=m-{\cal O}(T/{\cal Q})$.

As long as the net charge $\cal Q$ is a macroscopically large number
many orders of magnitude larger than $T/m$, this factor
coincides essentially with the condensate
contribution to the density $\rho_{0}$. One may gain intuition on how
charge is distributed among states by making a few
numerical exercises with our formulae. By way of example, we take a
fiducial volume of $10\, m^3$ filled with 400 units of
charge per cubic $cm$ (i.e. numerically equal to the photons in the
MWBR). Then, for $m=10^{-6}\,eV$, $4\times 10^{11}$
particles and 15 antiparticles populate the ground state, while
$4.6\times 10^{8}$ particles and $4.2\times 10^{8}$ fill the
excited states, at
$T=0.01\,T_{c}=3\times 10^{-4}eV$. Clearly, the statement following eq.
(23) is correct and it
allows us to calculate ${\cal T}(Q)\vert _{k=0}$ in terms of the fixed
quantity $\rho$ (eq.~(\ref{eq:rhotot})):
\be
{\cal T} (Q) \vert_{k=0}=\frac{i g^2}{3 Q^2} \left(T_c^2-T^2 \right).
\label{eq:mcondtc}
\ee
The Fourier transform of this equation gives the contribution of the
condensate to the potential. It is:
\be
V_0(r)=-\frac{g^2}{48 M^2 \pi} \frac{1}{r} \left(T_c^2-T^2 \right).
\label{eq:potcond}
\ee
The thermal contribution (i.e. from the excited states) is just
eq.~(\ref{eq:pother}) with the chemical
potential held fixed at the constant value $\mu=m$. The same expansion,
eq.~(\ref{eq:ident}), and the same integration techniques 
can be used now to generate the thermal potential below the condensation
temperature. Again, we find a piece that exactly
cancels the vacuum contribution to the interaction, and what finally
turns out to be the total potential, i.e.
the thermal contribution, plus the condensate contribution~(\ref{eq:potcond}), plus
the vacuum contribution~(\ref{eq:bessel}) is:
\be
V_{T\le T_c}^{total}=-\frac{g^2}{64 M^2 \pi^2}\left[{4\pi(T_c^2-T^2)\over
3r}+ \frac{1}{r^2}T\left(1+2\sum_{k=1}^\infty e^{-
 2rT\sqrt{2(\gamma +\eta)}}\cos
\left(2rT\sqrt{2(\gamma-\eta}\right)\right)\right]
\label{eq:vtotmenys}
\ee
where $\gamma\equiv k\pi\sqrt{k^2\pi^2+m^2/T^2}$ and $\eta\equiv
k^2\pi^2$. 

It is easy to check that the potential~(\ref{eq:vtotmenys}) for $T\le T_c$ and the
potential~(\ref{eq:vtotmes}) for $T\ge T_c$  coincide for $T=T_c$
($\xi=1$).

 Inspection of
these results leads us immediately to realize an important consequence
of
Bose-Einstein condensation. All terms in the infinite sum in eq.~(\ref{eq:vtotmenys})
decay faster than $e^{-2mr}$ for $T>m$.
 Therefore, for distances much larger than the Compton wavelength of
$\varphi$, i.e. $r\gg m^{-1}$ and hence  $r\gg
T_{c}^{-1}$, $V_{T\le T_c}^{total}$ is given by:
\be
V_{T\le T_c}^{total}\simeq -\frac{g^2}{48 M^2 \pi}
\frac{T_{c}^2}{r}\left[1+{\cal O}(\frac {T^2}{T_{c}^2},\frac
{T}{rT_{c}^2})\right].
\label{eq:vas}
\ee        
 Namely, at low
temperature (i.e. below $T_{c}$) the force, that was finite-ranged at
high temperature (i.e. above $T_{c}$), becomes infinite-ranged.
 This comes about because the medium absorbs and restores 3-momentum in
the
scattering process so that the 4-momentum squared
of the other $\varphi$ quantum exchanged in the $t$-channel
can reach the mass-shell in the physical region of the scattering
process. For the ${\mathbf k=0}$ mode in the bath, in particular, the
propagator of the second particle (see eq.~(\ref{eq:simpl}))
takes the form of the Coulomb propagator, and becomes singular 
at the edge of the Q-integration region, exactly as in the Coulomb
case. For
$T>T_{c}$ this infinite wavelength mode has zero measure, and it does not
contribute to the potential. However, in the
condensed phase, the infinite-range potential arises as a collective
phenomenon essentially because all charge piles up in
the ground state.

What we would like to do now is to show graphically the transition of
the potential as we vary the temperature from
$T>T_{c}$ to $T<T_{c}$. In the first temperature regime, i.e. above
condensation, we compute eq.~(\ref{eq:vtotmes}) numerically and below
$T_{c}$ we evaluate eq.~(\ref{eq:vtotmenys}) 
 where again numerical methods are used. 
Figure 2 displays our results.

\begin{figure}[bht]
\begin{center}
\epsfig{file=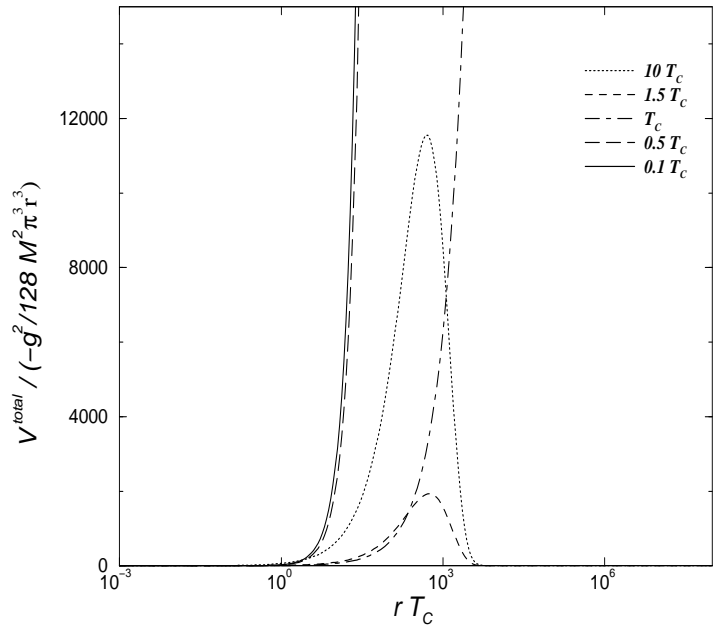,width=9.5cm,height=8.5cm}
\end{center}
\caption{\it Total potential, divided by $-g^2/128 M^2 \pi^3 r^3$, for
  a relativistic bose gas made of particles of $m \sim 10^{-6}eV$ and having a
  density $\rho \sim 40 \; cm^{-3}$. Behaviour above and below $T_c$ is
  shown. Note the Yukawa exponential damping for $T>T_c$ starting at $r \sim
  1/m \Rightarrow r\: T_c \sim 10^3$.}
\end{figure}

Let us briefly summarize our findings. Light scalars are basic
ingredients of many completions of the Standard model.
They may carry a new conserved quantum number. If ordinary matter is neutral
with respect to this new charge, then these scalars should couple
to ordinary matter pairwise. But double (pseudo)
scalar exchange generates long range spin independent forces among bulk
matter exactly as 2-neutrino exchange and
2-photon exchange (van der Waals forces) do. All these dispersion forces
are modified when matter is introduced in a heat
bath. The MWBR effect on van der Waals type potentials and the relic
neutrino cosmic
background effect on neutrino forces have been recently surveyed. The
present paper presents an investigation of the
effects of a relativistic ideal Bose gas on potentials generated by
2-scalar exchange. An example for such a heat bath
could be provided by hot dark matter (i.e relativistic at decoupling)
made of hypothetic relic light scalars. For this
purpose we use a very simple model for matter-scalar interactions for we
do not want to commit ourselves to any specific
extension of the SM. We do not expect that the particular $r$-behaviour
is in any sense realistic but the phenomena
produced by Bose-Einstein condensation of the heat reservoir are totally
independent of the form of the interaction chosen.
What we find is a very dramatic effect: below the critical temperature,
the finite-range force that we had above this
temperature becomes an infinite-range force. The phenomenon arises as a
combination of kinematics (3-momentum exchange of
the matter system with the medium) and the collective effect of
condensation of charge. In the particular model studied in
this paper, a potential of the form $\sim exp(-2mr)/r^2$ at $T>T_{c}$
converts to a $\sim 1/r$ potential at $T<T_{c}$.
Should hot relic scalars populate our Universe with a present density
such that their temperature is below the threshold
for Bose-Einstein condensation, then the effect described above would
provide an excellent opportunity for experiments
searching for forces weaker than gravity~\cite{fish} for in this case no
exponential
decay with distance occurs and, furthermore, a
milder power law fall-off with distance ensues.

Work partially supported by the CICYT Research Project
AEN98-1093. F.F. acknowledges the CIRIT for financial support.


\newpage
\begin{thebibliography}{99}
\bibitem{waals}
H. B. G. Casimir and P. Polder, Phys. Rev. {\bf 73} (1948) 360; E.M.
Lifschitz,
JETP  Lett. {\bf 2} (1956) 73; G. Feinberg and J. Sucher, J. Chem. Phys.
{\bf
  48} (1968) 3333; J. Soffer and J. Sucher, Phys. Rev. {\bf 161} (1967)
1664;
G. Feinberg and J. Sucher, Phys. Rev. {\bf A2} (1970) 2395.
\bibitem{neut}
G. Feinberg and J. Sucher, Phys. Rev. {\bf 166} (1968) 1638; J. A.
Grifols,
E. Masso and R. Toldra, Phys. Lett. {\bf B389} (1996) 363; E. Fischbach,
Ann. Phys. (N.Y.) {\bf 247} (1996) 213.
\bibitem{sik}
S. P. H. Hsu and P. Sikivie, Phys. Rev. {\bf D49} (1994) 4951.
\bibitem{pseudo1}
V.M. Mostepanenko and I.Yu. Sokolov, Sov. J. Nucl. Phys. {\bf 46} (1987)
685;
J. A. Grifols and S. Tortosa, Phys. Lett. {\bf B328} (1994) 98.
\bibitem{pseudo2}
F. Ferrer and J. A. Grifols, Phys. Rev. {\bf D58} (1998) 096006.
\bibitem{pseudo3}
F. Ferrer and M. Nowakowski, Phys. Rev. {\bf D59} (1999) 075009.
\bibitem{sucher}
G. Feinberg, J. Sucher and C.-K. Au, Phys. Rep. {\bf 180} (1989) 83;
G. Feinberg and J. Sucher, in {\it Long-Range Casimir Forces: Theory and
  Recent Experiments in Atomic Systems}, edited by Frank S. Levin and
David
A. Micha (Plenum, New York, 1993). 
\bibitem{fgcp}
F. Ferrer and J. A. Grifols, Phys. Lett. {\bf B460} (1999) 371.
\bibitem{fgn}
C. J. Horowitz and J. Pantaleone, Phys. Lett. {\bf B319} (1993) 186;
F. Ferrer, J. A. Grifols and M. Nowakowski, Phys. Lett. {\bf B446}
(1999) 111;
F. Ferrer, J. A. Grifols and M. Nowakowski, {\em hep-ph/9906463},
Phys. Rev. {\bf D} to be published.
\bibitem{consoli}
M. Consoli and P.M. Stevenson, {\em hep-ph/9711449}.
\bibitem{kobes}
See for instance, R. L. Kobes, G. W. Semenoff and N. Weiss,
Z. Phys. {\bf C29} (1985) 371.
\bibitem{haber}
H. E. Haber and H. A. Weldon, Phys. Rev. Lett. {\bf 46} (1981) 1497;
H. E. Haber and H. A. Weldon, Phys. Rev. {\bf D25} (1982) 502.
\bibitem{gradstein}
I. S. Gradshteyn, I. M. Ryzhik, {\it Table of Integrals, Series and Products},
corrected and enlarged edition (Academic Press, INC 1980).
\bibitem{fish}
E. Fischbach and C. L. Talmadge, {\it The Search for Non-newtonian
Gravity}
(Springer-Verlag, 1999).
\end{thebibliography}
\end{document}